\documentclass[11pt]{article}
\usepackage{graphicx}

\begin{document}

\def\xslash#1{{\rlap{$#1$}/}}
\def \p {\partial}
\def \dd {\psi_{u\bar dg}}
\def \ddp {\psi_{u\bar dgg}}
\def \pq {\psi_{u\bar d\bar uu}}
\def \jpsi {J/\psi}
\def \psip {\psi^\prime}
\def \to {\rightarrow}
\def\bfsig{\mbox{\boldmath$\sigma$}}
\def\DT{\mbox{\boldmath$\Delta_T $}}
\def\xit{\mbox{\boldmath$\xi_\perp $}}
\def \jpsi {J/\psi}
\def\bfej{\mbox{\boldmath$\varepsilon$}}
\def \t {\tilde}
\def\epn {\varepsilon}
\def \up {\uparrow}
\def \dn {\downarrow}
\def \da {\dagger}
\def \pn3 {\phi_{u\bar d g}}

\def \p4n {\phi_{u\bar d gg}}

\def \bx {\bar x}
\def \by {\bar y}


\begin{center}
{\Large\bf Evolution of Chirality-odd Twist-3 Fragmentation Functions }
\par\vskip20pt
J.P. Ma$^{1,2,3}$ and G.P. Zhang$^{4}$     \\
{\small {\it
$^1$ Institute of Theoretical Physics, Chinese Academy of Sciences,
P.O. Box 2735,
Beijing 100190, China\\
$^2$ School of Physical Sciences, University of Chinese Academy of Sciences, Beijing 100049, China\\
$^3$ Center for High-Energy Physics, Peking University, Beijing 100871, China\\
$^4$Department of Modern Physics,  University of Science and Technology of China, Hefei, Anhui 230026, China  
}} \\
\end{center}
\vskip 1cm
\begin{abstract}   
We derive the complete set of evolutions of chirality-odd twist-3 fragmentation functions at one-loop level.  
There are totally nine real twist-3 fragmentation functions, among which seven are independent. The renormalization-scale dependence of the nine functions has an important implication for studies of single transverse-spin asymmetries.   We find that the evolutions of the three complex fragmentation 
functions defined by quark-gluon-quark operator 
are mixed with themselves. There is no mixing with the fragmentation functions defined only with bilinear quark field operators. In the large-$N_c$ limit the evolutions of the three complex fragmentation 
functions are simplified and reduced to six homogenous equations.

\vskip 5mm
\noindent
\end{abstract}
\vskip 1cm

\par\vskip10pt
 
Fragmentation Functions(FF's) are important quantities in predictions made with QCD factorization for inclusive hadron production with large momentum transfers, whose generic scale is denoted as $Q$.   
Using QCD factorization, the production rate  can be  predicted as a convolution of  inclusive production rate 
of partons with FF's. The inclusive parton production rate can be calculated with perturbative theory of QCD.     
The produced partons will fragment into the observed hadron. Fragmentation processes are of long-distance effect and described 
by FF's. 

\par 
There are various FF's as ingredients of collinear- and transverse-momentum-Dependent factorization. 
The properties of FF's are recently reviewed in \cite{FFR}. In collinear factorization 
the most well-known FF's are of twist-2 defined with chirality-even QCD operators. These FF's appear in predictions of inclusive hadron production 
at the leading power of $Q^{-1}$, where no transverse polarization is observed.  
The renormalization-scale dependence of twist-2 FF's is governed by DGLAP-type evolutions and is known at two-loop level. 

\par 
If an initial hadron in Semi-Inclusive DIS and inclusive hadron production in hadron-collisions 
is transversely polarized, the production rate can contain transverse-spin dependent parts or single transverse-spin asymmetries can appear.  Such asymmetries are currently under intensive studies in theory and experiment. 
The asymmetries at the leading power of $Q^{-1}$ are predicted with twist-3 parton distributions and chirality-odd FF's at twist-3.   
Because of helicity conservation of perturbative QCD at high energy, contributions from chirality-odd FF's 
are always combined with the transversity, the parton distribution function of a transversely polarized hadron defined with a twist-2 chirality-odd operator in \cite{Trans}.  This distribution is less known than other 
standard twist-2 parton distributions. Therefore, studies of chirality-odd FF's are important not only 
for understanding  fragmentation and the asymmetries,  but also for extracting information about the transversity. 

\par 
Although fragmentation processes are in general nonperturbative and FF's at moment can only be extracted from experiment, 
but the renormalization-scale dependence can be calculated with perturbative QCD. It should be noted that 
this dependence will determine partly the $Q$-dependence of the transverse-spin-dependent part of the relevant 
differential cross-sections, similar to the case of DIS, where the scaling violation is predicted by the evolution 
of twist-2 parton distribution functions. Unlike the case of the evolution of FF's or PDF's defined with twist-2 operators, whose evolutions are now relatively easy to derive, there is a nontrivial problem in deriving evolutions of higher-twist operators. It is nontrivial to separate contributions at different twists as early studies of higher-twist effects in DIS have shown in \cite{TW4DIS,JWQ,G2}.  
In this letter, we study the renormalization-scale dependence of twist-3 chirality-odd FF's. Some of them have been studied in \cite{BeKu, Kang}. Here, we will derive all evolutions of them.

\par 
    
We notice that the same twist-3 operators used to defined chirality-odd FF's are also used to defined chirality-odd 
parton distributions. Because the time-reversal symmetry, the number of chirality-odd twist-3 parton distributions is smaller  than that of chirality-odd twist-3 FF's. The evolutions of chirality-odd twist-3 parton distributions have been studied in \cite{BBKT,KoNi,BD,Beli1,KQ2,MZGP}.  
The $Q$-dependence of single transverse-spin asymmetries are governed by the evolution of twist-3 parton distributions 
and twist-3 FF's combined with the evolution  of twist-2 operators. 
The evolution of chirality-even twist-3 parton distributions can be found in \cite{KQ2,KQ,BMP,VoYu,ZhSc,MW}. 
With these known evolutions and the evolution studied here, the $Q$-dependence can be predicted completely.    
Recently, the evolutions of twist-4 operators have been studied in\cite{JiBe}.

\par
For our purpose it is convenient to use the  light-cone coordinate system. In this system a
vector $a^\mu$ is expressed as $a^\mu = (a^+, a^-, \vec a_\perp) =
((a^0+a^3)/\sqrt{2}, (a^0-a^3)/\sqrt{2}, a^1, a^2)$ and $a_\perp^2
=(a^1)^2+(a^2)^2$. We introduce two light-cone vectors $n^\mu =(0,1,0,0)$ and $l^\mu =(1,0,0,0)$. 
The transverse metric in the coordinate system is given by $g_\perp^{\mu\nu} = g^{\mu\nu} - n^\mu l^\nu - n^\nu l^\mu$.  
With the metric the transverse part of any vector $a^\mu$ is obtained as $a_\perp^\mu = g_\perp^{\mu\nu} a_\nu$.

\par 
The definitions of twist-3 chirality-odd FF's have been discussed in \cite{JiFF,MPF,EKT}. We assume that  the produced hadron through parton fragmentations moves in the $+$-direction with the momentum $P^\mu =(P^+,0,0,0)= P^+ l^\mu$.  We will work in the light-cone gauge $n\cdot G= G^+ =0$. In this gauge  gauge links, which are needed in other gauges to make 
the definitions gauge-invariant, are always a unit matrix. From two-parton correlation functions one can define  three chirality-odd FF's for the unpolarized hadron in $d$-dimensional space-time:     
\begin{eqnarray} 
  \hat e(z) &=& \frac{z^{d-3} P^+ }{4 N_c }\int \frac{d\lambda}{2\pi} e^{-i\lambda P^+/z }
   \sum_X  {\rm Tr} \langle 0\vert  \psi (0) \vert h X\rangle    \langle h X \vert \bar \psi (\lambda n )   \vert 0 \rangle, 
\nonumber\\
  \hat e_I (z) &=& \frac{z^{d-3} P^+ }{4 N_c }\int \frac{d\lambda}{2\pi} e^{-i\lambda P^+/z }
   \sum_X  {\rm Tr} \langle 0\vert  \sigma^{-+} \psi (0) \vert h X\rangle    \langle h X \vert \bar \psi (\lambda n )   \vert 0 \rangle,
\nonumber\\    
\hat e_\partial (z) &=& \frac{z^{d-3}}{4 N_c (d-2) }   \int \frac{d\lambda}{2\pi} e^{-i\lambda P^+/z }
   \sum_X  {\rm Tr} \langle 0\vert \gamma^+ \gamma_{\perp\mu}  (0) \psi(0) \vert h X\rangle \partial_\perp^\mu    \langle h X \vert \bar \psi (\lambda n )   \vert 0 \rangle. 
\label{2PFF}             
\end{eqnarray}     
The defined three functions are real and have the support of $ 0\le z \le 1$ for fragmentation of a quark. 
In the case of anti-quark FF's one has $ -1 \le z \le 0 $. 
Since the number of $\gamma$-matrices sandwiched between two quark fields in Eq.(\ref{2PFF}) is even, these fields must have different chiralities, or different helicities in the massless case. If chiral-symmetry is an exact symmetry of QCD, these functions are zero. It is noted that time-reversal symmetry of QCD can not be used here to obtain constraints on these functions.
In Eq.(\ref{2PFF}) the state $\vert h X\rangle$ is an out-state at the time $t=\infty$. Under the transformation of time-reversal, the state becomes an in-state at the time $t=-\infty$. There is in general no simple relation between the in- and out states. If there are no final- or initial state interactions, the out-state can be related to the in-state through a 
phase factor, or the amplitude $\langle h X\vert \bar \psi \vert 0\rangle$ contains only dispersive part and no absorptive part. In this case, one can use the time-reversal symmetry to show that $\hat e_I$ and $\hat e_\partial$ are zero. 
But final-state- or initial-state interactions always exist, the absorptive part of the amplitude $\langle h X\vert \bar \psi \vert 0\rangle$ can be nonzero. Therefore, $\hat e_I$ and $\hat e_\partial$ are not zero in general.


\par
From the three-parton correlations one can define three complex FF's: 
\begin{eqnarray} 
 \hat E_F (z_1,z_2) &=&  - \frac{z_2 g_s}{2(d-2) N_c}  \int \frac{d\lambda_1 d\lambda_2}{(2\pi)^2} e^{-i\lambda_1 P^+/z_1 -i\lambda_2P^+/{z_3} }  
\nonumber\\ 
  && \sum_X {\rm Tr} \langle 0\vert i \gamma^+ \gamma_{\perp\mu}  \psi(0) \vert h X\rangle \langle h X \vert \bar \psi (\lambda_1 n ) G^{+\mu}(\lambda_2 n) 
   \vert 0 \rangle ,
\nonumber\\
   \hat E_{\bar F}  (z_1,z_2) &=&  \frac{z_2 g_s}{4 N_c} \frac{2}{d-2}  \int \frac{d\lambda_1 d\lambda_2}{(2\pi)^2} e^{-i\lambda_1 P^+/z_1 -i\lambda_2P^+/{z_3} }  
\nonumber\\ 
  &&  \sum_X {\rm Tr} \langle 0\vert \bar \psi (0) ( i \gamma^+ \gamma_{\perp\mu})   \vert h X\rangle\langle h X \vert  G^{+\mu}(\lambda_2 n) \psi (\lambda_1 n )
   \vert 0 \rangle ,   
\nonumber\\
   \hat E_G (z_1,z_2) &=& -\frac{z_2 g_s}{4 (N_c^2-1) } \frac{2}{d-2} \int \frac{d\lambda_1 d\lambda_2}{(2\pi)^2} e^{i \lambda_1 P^+/z_1 -i\lambda_2P^+/z_2  }
\nonumber\\  
   && \sum_X {\rm Tr} \langle 0\vert \bar \psi (\lambda_1 n ) i \gamma^+ \gamma_{\perp\mu} T^a \psi(0) \vert h X\rangle \langle h X \vert  G^{a, +\mu}(\lambda_2 n) 
   \vert 0 \rangle
\label{3PFF}     
\end{eqnarray}
with $1/{z_3} = 1/z_2 -1/z_1$. $\hat E_{\bar F}$ is for fragmentation with an anti-quark.  
In general the three functions are complex. If there is no final state interaction, one can use time-reversal symmetry, as 
discussed before, to show that they are real. The functions have the support: 
\begin{eqnarray} 
  0 < z_2 < 1,\quad {\rm or } \quad  z_2 < z_1 <\infty.   
\end{eqnarray}   
In \cite{MeMe} it is shown that these functions are zero at $z_1=z_2$ or  $1/z_1=0$. 

\par    
The introduced FF's are defined with chirality-odd QCD operators of twist-3.  They are not independent. From QCD equation of motion one can derive the relations\cite{BeKu,MPF,EKT}: 
\begin{eqnarray}
 2 z_2 \hat e_\partial (z_2) - \hat e_I(z_2) = z_2^2 \int \frac{d z_1}{z_1} P\frac{1}{ z_2-z_1} {\rm Im } \hat E_F (z_1,z_2), 
 \quad \hat e(z_2) = z_2^2 \int \frac{d z_1}{z_1} P\frac{1}{z_2-z_1} {\rm Re }\hat E_F(z_1,z_2).  
\label{RL2N}
\end{eqnarray}
With these relations we can take $\hat e$ and $\hat e_I$ as redundant. This is a reasonable choice as we will explain later.
It is interesting to compare the defined FF's with parton distributions defined with the same chirality-odd operators. Taking $\hat E_F$ as an example, one can define a parton 
distribution with the same operator in $\hat E_F$ as its matrix element of a single hadron. Because the state is of a single stable hadron, the in- and out state are related through a phase factor. One can use time-reversal symmetry of QCD to show that  the defined parton distribution is real. For $\hat E_F$ itself, the symmetry does not give any constraint. This results in that $\hat E_F$ is in general a complex function.   
 
\par  
The defined FF's in Eq.(\ref{2PFF},\ref{3PFF})  depend not only on momentum fractions, but also on the renormalization scale $\mu$ implicitly. The $\mu$-dependence or evolution of the FF's is the subject of our study. 
In \cite{BeKu} the evolutions of the real parts of $\hat E_{F,G}$ have been studied.  The evolution 
of $\hat e_\partial$ is obtained in \cite{Kang}. In general different FF's can be mixed under evolutions. 
We can divide FF's into two groups. One group contains $\hat e$ and the real part of $\hat E_{F,\bar F, G}$. These FF's can be called as $T$-even, while another group contains $\hat e_I$, $\hat e_\partial$ and the imaginary parts of $\hat E_{F,\bar F, G}$. The FF's in this group can be called as $T$-odd because they are 
zero in the absence of final-state interactions. Under the evolution two groups of FF's can not be mixed with each other, 
because the evolution kernels do not contain absorptive parts, at least at one-loop level.   
For $T$-even FF's we may only need to determine the evolution of ${\rm Re}\hat E_{F,\bar F,G}$. The evolution of $\hat e$ 
may be determined by using the second relation in Eq.(\ref{RL2N}). For $T$-odd FF's, one may determine the evolution 
of $\hat e_I$ with the first relation in Eq.(\ref{RL2N}), once one knows the evolution of ${\rm Im}\hat E_F$ and $\hat e_\partial$. In this letter, we will 
directly derive the evolution of each FF and use the relations to check our results.

\par 
\begin{figure}[hbt]
\begin{center}
\includegraphics[width=12cm]{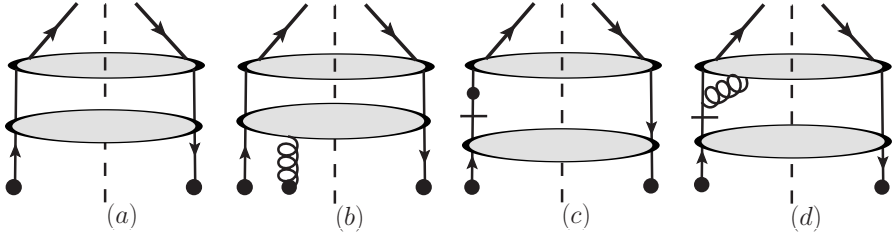}
\end{center}
\caption{The structure of diagrams for evolutions.  } 
\end{figure}
\par 
To derive the evolutions of FF's, we essentially need to study, e.g., contributions of Fig.1a and Fig.1b. 
Fig.1a and Fig.1b represent the generic structure 
of diagrams for the contributions from two-parton FF's to the evolution of 
two-parton- and three-parton FF's, respectively. The black dots there denote the insertion of the field operators
with given projections of $\gamma$-matrices in the definitions of FF's.  The middle bubbles represent diagrams of parton scattering. The top bubbles
denote the two-parton density matrix $\Gamma$, which is defined and can be decomposed as: 
\begin{eqnarray} 
  \Gamma_{ji} (q) &=& \int \frac{d^d \xi}{(2\pi)^d} e^{-i \xi\cdot q} \sum_X \langle  0\vert \psi_j (0)  \vert P X \rangle  \langle  X P \vert \bar \psi_i (\xi) \vert 0 \rangle 
 \nonumber\\
      &=&    \frac{\delta (q^-) }{ z^{d-3} P^+} \biggr [ \delta^{d-2} (q_\perp) \biggr ( \gamma^- P^+ d(z)+  \hat e (z) +  \sigma^{+-}  \hat e_I(z)   \biggr )   - i \gamma^- \gamma_\perp^\mu  P^+ \hat e_\partial (z) \frac{\partial}{\partial q_\perp^\mu }\delta^{d-2} (q_\perp)\biggr ]_{ji} 
   +\cdots, 
 \label{Ga2}        
\end{eqnarray}    
where $\cdots$ denote terms at twist higher than $3$. The plus component of $q$ is fixed as $q^+ =P^+/z$.
$d(z)$ is the standard twist-2 FF defined in \cite{PDFFF}. Other three functions are those FF's defined in Eq.(\ref{2PFF}). When we take the middle part at the tree-level,  i.e., at order of $\alpha_s^0$, or delete the middle bubbles, 
from Fig.1a one simply obtains corresponding two-parton FF's themselves, and from Fig.1b the three-parton FF's receive no contributions from two-parton FF's. When the middle part is at one-loop level or at order of $\alpha_s$, there are U.V. divergences. Renormalization is hence needed and the renormalization-scale dependence appears. This dependence determines 
the evolutions.    
\par 
As mentioned at the beginning, there is a nontrivial problem   
 of how to consistently separate contributions at different twists. The problem  can be explained with Fig.1a or Fig.1b. 
With the decomposition of $\Gamma$ in Eq.(\ref{Ga2}) one can see that Fig.1a or Fig.1b contains the contributions at twist-2. 
However the above decomposition can not be used here for finding the twist-3 contributions, because the so-called bad components of quark fields 
in $\Gamma$ appear for twist-3 contributions. They appear in the definition of $\hat e$ and $\hat e_I$. 
The bad component is defined by decomposing the quark field $\psi$ as:       
\begin{equation}
 \psi (x) =\psi_+ (x) + \psi_-(x), \quad  \psi_-(x) = \frac{1}{2} \gamma^+ \gamma^- \psi (x), \quad \psi_+(x)  = \frac{1}{2} \gamma^- \gamma^+ \psi(x),   
\end{equation}
where $\psi_-$ is called as the bad component and $\psi_+$ is the good component. 
It is easy to realize that the field $\psi_-$ can not be interpreted as a field for creating- or annihilating 
a quark moving in the $+$- direction. The propagator containing $\psi_-$ does not propagate. 
 Therefore, $\psi_-$ can not be used with perturbation theory here directly. That 
is the reason for the un-efficiency of using the decomposition to separate the contributions at different twists, especially, 
beyond tree-level. 
\par 
The solution of the problem is to express the bad component with the good component combined with gluon fields through 
equation of motion in the light-cone gauge:  
\begin{equation} 
  2 \partial^+ \psi_-  + \gamma^+ \gamma_\perp\cdot D_\perp \psi_+ =0, \quad   2 D^- \psi_+  + \gamma^- \gamma_\perp\cdot D_\perp \psi_- =0 
\label{EQGB}   
\end{equation}
with the covariant derivative $D^\mu = \partial^\mu + i g_s G^\mu$. 
By using $\psi_\pm$ the operator $\psi(0) \bar \psi (\xi)$ in $\Gamma$ can be written 
with various combinations of $\psi_{\pm}$ and these combinations give the different parts in the second line of Eq.(\ref{Ga2}): 
\begin{eqnarray} 
  \Gamma (q)  &=& \Gamma^{ (++) } (q) +  \Gamma^{(+-)}  (q)  +\Gamma^{(+-)\dagger }  (q)  +\Gamma^{(--)} (q), 
\nonumber\\
   \Gamma^{ (++) } (q) &=&  
     \int \frac{d^d \xi}{(2\pi)^d} e^{-i \xi\cdot q} \sum_X \langle  0\vert \psi_{+} (0)  \vert P X \rangle  \langle  X P \vert \bar \psi_{+} (\xi) \vert 0 \rangle
\nonumber\\
        &=&  \frac{\delta (q^-) }{ z^{d-3} P^+} \biggr ( \delta^{d-2} (q_\perp)  \gamma^- P^+ d(z)    - i \gamma^- \gamma_\perp^\mu  P^+ \hat e_\partial (z) \frac{\partial}{\partial q_\perp^\mu }\delta^{d-2} (q_\perp)\biggr ) 
  +\cdots,         
\nonumber\\
   \Gamma^{(+-)}  (q)  &=&  \int \frac{d^d \xi}{(2\pi)^d} e^{-i \xi\cdot q} \sum_X \langle  0\vert \psi_{+} (0)  \vert P X \rangle  \langle  X P \vert \bar \psi_{-} (\xi) \vert 0 \rangle, 
\nonumber\\     
\Gamma^{(+-)}  (q) +  \left ( \Gamma^{(+-)} (q)  \right )^\dagger  &=& \frac{\delta (q^-) }{ z^{d-3} P^+} \delta^{d-2} (q_\perp) \biggr (   \hat e (z) +  \sigma^{+-}  \hat e_I(z)   \biggr ) 
   +\cdots.   
\end{eqnarray}  
$\Gamma^{(--)}$ stands for the combination with $\psi_-(0) \bar \psi_-(\xi)$ and is of twist higher than 3.  
\par 
We use the equation of motion in Eq.(\ref{EQGB}) to express $\bar \psi_-$ in $\Gamma^{(+-)}$ with $\bar\psi_+$: 
\begin{eqnarray} 
  \Gamma^{(+- )}  (q) &=&   \biggr [ \int \frac{d^d \xi}{(2\pi)^d} e^{-i \xi\cdot q} \sum_X \langle  0\vert \psi_{+} (0)  \vert P X \rangle   \langle  X P \vert  \partial_\perp^\mu\bar \psi_+ (\xi)  \vert 0 \rangle  \biggr ]  
   \gamma_{\perp\mu} 
   \frac{i \gamma^+}{ 2 q^+}   + \int d^d k \biggr [ \int \frac{d^d \xi d^d \xi_g}{ (2\pi)^{2d} }
 \nonumber\\
  &&   e^{-i\xi\cdot k -i \xi_g \cdot  (q-k) }  \sum_X \langle  0\vert \psi_{+} (0)  \vert P X \rangle
  \langle  X P \vert \bar \psi_+ (\xi) G_\perp^{a,\mu} (\xi_g)    \vert 0 \rangle \biggr ]   (-i g_s \gamma_\mu T^a ) \frac{ i \gamma^+}{2 q^+}, 
\label{GPM}    
 \end{eqnarray}  
where irrelevant terms are neglected. In Eq.(\ref{GPM}) the first term can be expressed by Fig.1c, where the top-bubble 
represents the correlation function in $[\cdots]$ in the first term. The black dot near the top bubble denotes 
the vertex $\gamma_{\perp}^\mu$.  
The second term can be expressed with Fig.1d, where the top-bubble 
represents the quark-gluon correlation function in $[\cdots]$. In Fig.1c and 1d the quark line with the short bar denotes the special propagator $ i\gamma^+/(2 q^+)$. The propagator is just the contraction of $\bar \psi_-$ with $\psi$, it has 
no pole at $q^2 =0$\cite{JWQ}.      
\par 
With the above discussion it is now clear how to correctly separate contributions at different twists from Fig.1a. 
The twist-3 contributions from  2-parton FF's  can be obtained by calculating Fig.1a with the top-bubble represented 
by $\Gamma^{(++)}$. The contributions from Fig.1c and Fig.1d have to be included with the rules discussed in the above. 
In Fig.1a, 1b and 1c, the top-bubbles only contain good components of quark fields. The contributions from Fig.1b are calculated in the similar way. There are contributions with an additional gluon-line connecting two bubbles. Their 
leading contributions are already at twist-3 and have no bad components in the top bubble. Since we will only deal with 
diagrams whose top bubbles contain good components of quark fields only, it is natural to take 
$\hat e$ and $\hat e_I$ as redundant in evolutions.      

\par 
\begin{figure}[hbt]
\begin{center}
\includegraphics[width=14cm]{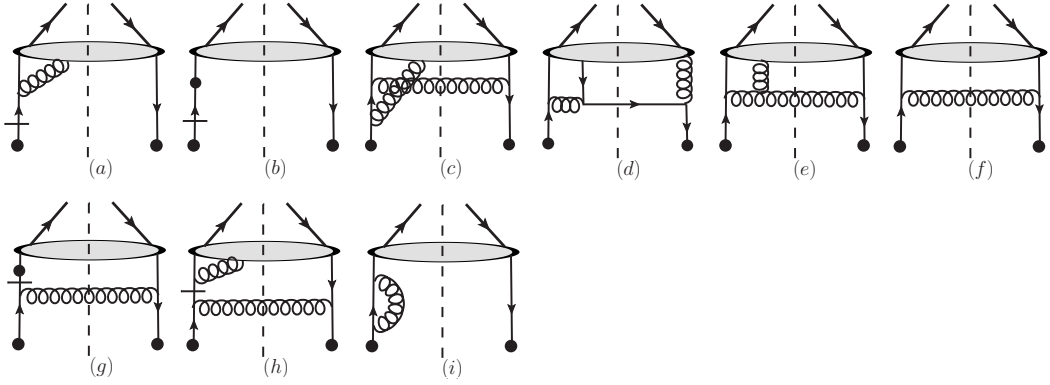}
\end{center}
\caption{Diagrams for the evolution of 2-parton FF's.   } 
\end{figure}

\par 
We first turn to the evolutions of 2-parton FF's. At one-loop we need to study contributions from diagrams given in Fig.2. 
A direct calculation of Fig.2a and 2b gives the results as the relations in Eq.(\ref{RL2N}) with $\hat e_\partial =0$. There 
is no contribution to $\hat e_\partial$ from Fig.2a and 2b. Inserting one-loop correction to propagators and vertices into Fig.2a and Fig.2b, 
we obtain the explicit renormalization-scale dependence, hence the contributions to evolutions. The contributions to  evolutions from the remaining diagrams can be calculated directly. In calculations all contributions 
are proportional to the integral 
\begin{equation} 
 \mu^{\epsilon} \int \frac{d^{d-2} k_\perp}{(2\pi)^{d-2}} \frac{1}{k_\perp^2}=\frac{1}{4\pi} 
   \biggr ( \frac{2}{\epsilon} -\gamma + \ln (4\pi)  -\frac{2}{\epsilon_c} + \ln\frac{e^\gamma \mu^2}{4\pi \mu_c^2}\biggr ),  
\end{equation}     
which is regularized with the dimensional regularization in $d=4-\epsilon$ space-time. The first pole of $1/\epsilon$ is from 
the U.V. divergence, which is subtracted by renormalization. $\mu$ is the renormalization scale. The pole in $\epsilon_c=4-d$ 
is from collinear divergence and $\mu_c$ is the related scale. The contributions from Fig.2a and Fig.2b by inserting  one-loop correction to propagators and vertices, and those from Fig.2i can be called as the virtual part, because 
there is no parton crossing the cut. The contributions from the remaining diagrams are called as the real part, where there is one gluon crossing the cut. 
\par 
In calculating the virtual part, we find that Fig.2b and Fig.2i give no contribution to the evolution of  $\hat e$. Fig.2a and Fig.2b give 
no contribution to the evolution of $\hat e_\partial$ because of $\gamma^+ \gamma^+ =0$. 
We have the total virtual part of the evolutions of two-parton FF's:
\begin{eqnarray} 
  \mu \frac{\partial \hat e (z) }{\partial \mu}\biggr\vert_{vir.} &=&  
      \int \frac{ d x_1 }{x_1} {\mathcal K}_v (z, x_1) \frac{z^2}{z-x_1}  {\rm Re} \hat E_F ( x_1, z),
\nonumber\\ 
     \mu \frac{\partial \hat e_I (z) }{\partial \mu}\biggr\vert_{vir.}  &=&  -  \int \frac{ d x_1 }{x_1} 
      {\mathcal K}_v (z, x_1) \frac{z^2}{z-x_1}  {\rm Im} \hat E_F ( x_1, z) +\frac{\alpha_s C_F}{\pi}     \biggr ( 3 + 4 \ln z - 4 \int_0^1 \frac{dy }{y} \biggr ) \hat e_\partial (z),
\nonumber\\ 
     \mu \frac{\partial \hat e_\partial (z) }{\partial \mu}\biggr\vert_{vir.}  &=& \frac{\alpha_s C_F}{2\pi}     \biggr ( 3 + 4 \ln z - 4 \int_0^1 \frac{dy }{y} \biggr ) \hat e_\partial (z),     
\label{2PVIR}                      
\end{eqnarray}  
with 
\begin{eqnarray}       
   {\mathcal K}_v(z, x_1)    &=& \frac{\alpha_s}{\pi} 
   \biggr \{  C_F \biggr ( \frac{1}{2} + \ln (z x_1) - 2 \int_0^1 \frac{dy}{y}   \biggr )
 +     
    \frac{1} {2 N_c}  \frac{z }{x_1-z} \ln \frac{z}{x_1}  
      + \frac{N_c}{2}   
         \ln \frac{z}{x_1}  \biggr \}. 
\end{eqnarray} 
The above results contain a divergence in the integral of $y$ as the momentum fraction in unit of $P^+$. 
In the light-cone gauge the gluon propagator becomes:
\begin{equation}
    \frac {i}{ k^2 +i\varepsilon} \biggr ( - g^{\mu\nu} +\frac{ n^\mu k^\nu + k^\mu n^\nu}{n\cdot k} \biggr ). 
\end{equation}      
The divergence comes from the part with $1/n\cdot k$. This divergence is the so-called light-cone divergence. 
It will be cancelled in the final results.  

\par 
The calculation of the real part is straightforward. Adding the real part to the virtual part we have the 
evolutions of two-parton FF's at one-loop:
\begin{eqnarray} 
 \mu\frac{\partial \hat e(z, \mu)}{\partial \mu} &=& \frac{z \alpha_s}{\pi} \int \frac{ d\xi_1 d \xi_2}{\xi_1 \xi_2} 
          \biggr \{ {\rm Re}\hat E_F (x_1,x_2) \biggr [  
      - C_F   \frac{2 \xi_1\xi_2}{(1-\xi_2)_+ (\xi_2-\xi_1)} 
 + \frac{N_c} {2 }  \frac{\xi_1}{\xi_2-\xi_1} + \frac{1}{2 N_c} 
\nonumber\\     
      &&    -  \frac{\xi_1\delta (\xi_2-1)}{2 (1-\xi_1) } \biggr (  C_F  
            + \frac{ \ln \xi_1}{ N_c (1-\xi_1)} \biggr )   \biggr ]   
   +\frac{C_F } { N_c}   {\rm Re} \hat E_G (x_1,x_2) \frac{1-\xi_2 }{1-\xi_2+\xi_1} \biggr \}, 
\nonumber\\
     \mu \frac{\partial \hat e_\partial  (z,\mu)}{\partial \mu } & =& \frac{\alpha_s C_F }{\pi}\biggr [ \frac{3}{2}  \hat e_\partial (z) +  2  \int_z^1\frac{d\xi}{\xi}   \frac{1}{(1-\xi)_+} \hat e_\partial (x) \biggr ]  
  + \frac{\alpha_s}{\pi} \int \frac{d \xi_1 d \xi_2}{\xi_1 \xi_2} \biggr \{ {\rm Im } \hat E_F (x_1,x_2)      
\nonumber\\
   &&   \biggr [   \frac{N_c\xi_2}{2 (\xi_2-\xi_1)^2} \biggr (-\xi_1 -\frac{\xi_2(1-\xi_2)}{1-\xi_1} \biggr )   - C_F \biggr ( \xi_2 + \frac{ \xi_2}{\xi_1-\xi_2}\biggr ) \biggr ] 
\nonumber\\   
  && - \frac{C_F}{ N_c} {\rm Im } \hat E_G (x_1,x_2)\frac{ (1-\xi_2)^2}{1-\xi_2+\xi_1}   \biggr ] \biggr \} , 
\nonumber\\  
    \mu \frac{\partial \hat e_I  (z,\mu)}{\partial \mu } & =& \frac{\alpha_s }{\pi}\biggr \{ 3 z C_F \hat e_\partial (z) +  4z  C_F \int_z^1\frac{d\xi}{\xi}   \frac{1}{(1-\xi)_+} \hat e_\partial (x)  
    \biggr \}  + \frac{z \alpha_s}{\pi} \int \frac{ d\xi_1 d \xi_2}{\xi_1 \xi_2} 
          \biggr \{ {\rm Im }\hat E_F (x_1,x_2)
\nonumber\\
   &&  \biggr [  
       C_F   \frac{ 2 \xi_1 (\xi_2^2-\xi_1 ) }{(1-\xi_2)_+ (\xi_2-\xi_1)^2} 
 - \frac{1} {2N_c }  \frac{\xi_2 (\xi_1 +\xi_2) }{(\xi_2-\xi_1 )^2} +   \frac{\xi_1\delta (\xi_2-1)}{2 (1-\xi_1) } \biggr (  C_F  
            + \frac{ \ln \xi_1}{ N_c (1-\xi_1)} \biggr )  
\nonumber\\     
      &&   - C_F \frac{\xi_1}{\xi_2-\xi_1}     \biggr ]   
 -\frac{C_F } { N_c}   {\rm Im} \hat E_G (x_1,x_2) \frac{1-\xi_2 }{1-\xi_2+\xi_1} \biggr \},     
\label{2PEVO}                        
\end{eqnarray} 
with the notation and the integration measure:
\begin{equation} 
   \xi=\frac{z}{x}, \quad \xi_{1,2} =\frac{z}{x_{1,2}}, \quad \int \frac{ d\xi_1 d \xi_2}{\xi_1 \xi_2} = \int_z^1 \frac{d\xi_2}{\xi_2} \int_0^{\xi_2}\frac{ d\xi_1}{\xi_1} 
       = \int_z^1 \frac{ d x_2}{x_2} \int_{x_2}^\infty \frac{ d x_1}{x_1}. 
\end{equation}        
In the above results the light-cone singularities are cancelled. After the cancellation the standard $+$-distribution appears, 
which is defined as:  
\begin{eqnarray}
   \int_0^1 d z \frac{\theta (z-x)}{(1-z)_+} f(z) = \int_x^1 dz \frac{f(z)-f(1)}{1-z} + f(1) \ln (1-x). 
\end{eqnarray}

\par 
\begin{figure}[hbt]
\begin{center}
\includegraphics[width=14cm]{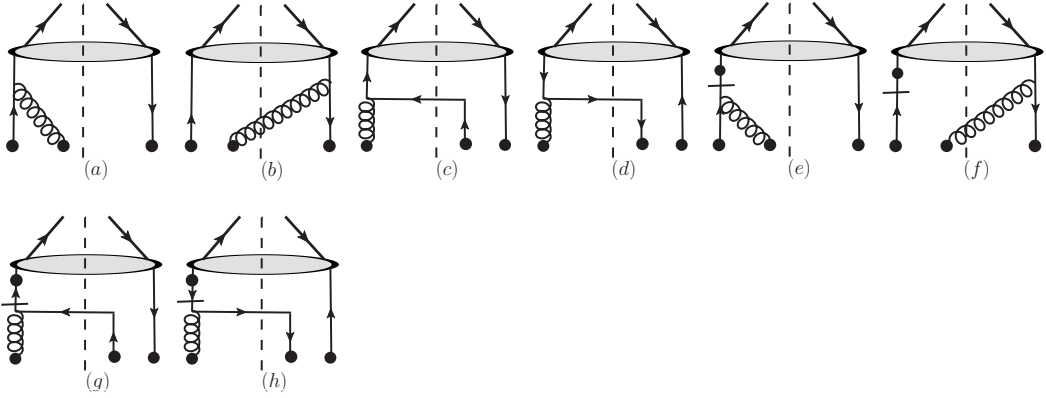}
\end{center}
\caption{Diagrams for the evolution of three-parton FF's.   } 
\end{figure}

\par 
Now we turn to the evolution of three-parton FF's. There are contributions from two-parton FF's to the evolution 
given by Fig.3. Interestingly, the sum of the contributions is zero. E.g., from Fig.3a and Fig.3e   
\begin{eqnarray}
 \mu\frac{\partial\hat E_F(z_1,z_2) }{\partial \mu } \biggr\vert_{3a} &=& 2 i z_2 \frac{\alpha_s C_F} {\pi} \frac{z_1-z_2}{z_1^2 }  \hat e_\partial (z_2),   
\nonumber\\ 
 \mu\frac{\partial\hat E_F(z_1,z_2) }{\partial \mu } \biggr\vert_{3e} &=& -2 i z_2 \frac{\alpha_s C_F} {2\pi} \frac{z_1-z_2}{z_1^2 }  \hat e_\partial (z_2). 
\end{eqnarray} 
The sum of the two diagrams is zero. The cancellation for other diagrams happens in a similar way. Therefore, 
the evolution of $\hat E_{F,G}$ only involves three-parton FF's. There is no mixing with two-parton FF's at one-loop.

\par 
The virtual part of the evolutions of $\hat E_{F,G}$ is simple. It is from the quark- or gluon self-energy correction, and the $\mu$-dependence of $g_s$ in the definitions of $\hat E_{F,G}$.  
We have: 
\begin{eqnarray} 
   \mu\frac{\partial\hat E_F(z_1,z_2) }{\partial \mu } \biggr\vert_{vir.}     &=&  
     \frac{\alpha_s }{\pi}   \biggr [ 
       C_F \biggr (\frac{3}{2}  +  \ln z_1 z_2  -2 \int _0^{1 }\frac{d y }{y} \biggr ) 
   +  
    N_c \biggr (  \ln z_3  -\int_0^{1} \frac{dy}{y} \biggr ) \biggr ] \hat E_F (z_1,z_2), 
\nonumber\\
   \mu\frac{\partial\hat E_G (z_1,z_2) }{\partial \mu } \biggr\vert_{vir.}     &=&  
     \frac{\alpha_s }{\pi}  \biggr [ 
       C_F \biggr (\frac{3}{2}  +  \ln z_1 z_3  -2 \int _0^{1 }\frac{d y }{y} \biggr ) 
   +  
    N_c \biggr (  \ln z_2  -\int_0^{1} \frac{dy}{y} \biggr ) \biggr ]  \hat E_G (z_1,z_2). 
\label{VIEF}            
\end{eqnarray}
Again, in this part there are light-cone singularities.

\par

\par
\begin{figure}[hbt]
\begin{center}
\includegraphics[width=16cm]{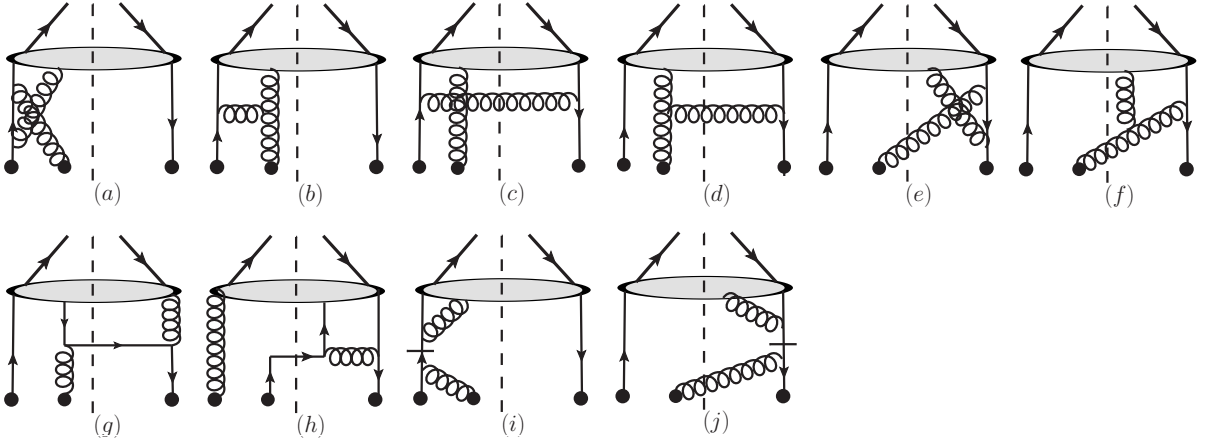}
\end{center}
\caption{ Diagrams for the $\hat E_F$-evolution.  } 
\end{figure}

\par  
The real part of the evolution of $\hat E_F$ or $\hat E_G$ is from Fig.4 or Fig.5, respectively. The calculations 
of these diagrams are straightforward. Adding the real part, the light-cone singularity in the virtual part will be 
cancelled. We introduce the following notations for giving our results: 
\begin{equation} 
 \frac{1}{x_{13}} = \frac{1}{x_1} + \frac{1}{z_3},\quad \frac{1}{x_{11}} = \frac{1}{x_1} + \frac{1}{z_1},\quad \xi_{1}=\frac{z_{1}}{x_{1}},  \quad \xi_{2}=\frac{z_{2}}{x_{2}} .
\end{equation} 
The evolution of $\hat E_F$ reads: 
\begin{eqnarray} 
   \mu \frac{\partial \hat E_F (z_1,z_2) }{\partial \mu }  &=& \frac{\alpha_s }{\pi} \biggr \{ 
     \biggr ( C_F \biggr ( \frac{3}{2} + \ln \frac{z_2}{z_1} \biggr ) + N_c \ln \frac{ z_3}{z_2} \biggr ) \hat E_F (z_1,z_2)
     + N_c \int_{z_2/z_1}^1 \frac{ d\xi_2}{(1-\xi_2)_+} \hat E_F(z_1,x_2) 
\nonumber\\
    &&  +  \int_{0}^1  \frac{d\xi_1} {(1-\xi_1)_+}  \biggr ( N_c \hat E_F (x_1,z_2) - \frac{1}{N_c} \hat E_F (x_1,x_{13}) \biggr )  
  - \frac{N_c}{2} \int^{z_1/z_2}_0 \frac{ d\xi_1}{1-\xi_1} \hat E_F(x_1,z_2) 
\nonumber\\      
   && + \frac{1}{ z_3 } \int  \frac{d x_1}{x_1} \biggr \{\hat E_F(x_1,z_2)   \biggr [ \frac{1}{2 N_c} 
 \frac{ z_2^2 (x_1-z_3)}{ z_1   (x_1-z_2)}
  \biggr ( \theta (z_3-x_1) \frac{z_1}{z_3}   
  + \theta (x_1-z_3) \frac{z_2}{x_1-z_2} \biggr )
\nonumber\\
   &&  + \frac{N_c}{2} 
   \biggr (  \theta (x_1-z_1)  \frac{ z_2^2 (-x_1 z_2-x_1 z_3 +z_2 z_3)}{z_1 (x_1-z_2)^2}     
  - \theta (z_1-x_1)\frac{z_2^2 (x_1-z_3) }{z_3 (x_1-z_2) }     \biggr ) 
\nonumber\\
    && - C_F \frac{z_2^3}{z_1 (x_1-z_1) } \biggr ]   -\frac{1 } {2 N_c }  \hat E_F(x_1,x_{13})  z_2 \theta (x_1-z_1)    
 \nonumber\\  
 &&
 + \frac{1} {2} \hat E_F^*(x_1,z_1) \frac{z_1z_2}{z_3(x_1-z_1)} \biggr [  2 C_F (x_1+z_3)   + N_c \frac{ x_1^3}{(x_1-z_2)(x_1-z_1) }
  \biggr ] \biggr \}   
\nonumber\\
   && + \frac{ N_c } {2 }  \int  \frac{d x_2}{x_2} \hat E_F(z_1,x_2)
      \theta(x_2-z_2) \biggr ( -\frac{z_2}{x_2} + \frac{z_1 ( z_1 x_2 + z_3 (z_1-x_2))}{z_3(x_2-z_1)^2 } \biggr )               
\nonumber\\
    && -\frac{ C_F } { N_c } \int \frac{d x_1}{x_1} \biggr [  \hat E_G^*( x_1,  x_{11})\frac{ (x_{11}-z_2)^2 }{ x_{11} z_2}\theta (x_1-z_3) 
 +  \hat E_G (x_1,z_3)
     \frac{ x_1 z_2  }{ z_3 (x_1+z_1 )} \biggr ] \biggr \} .  
\label{EVOEF}                       
\end{eqnarray}
The principal description of the last integral in the second line in Eq.(\ref{EVOEF}) is implied.  
The evolution of $\hat E_F$ involves not only $\hat E_F$ and $\hat E_G$, but also their complex conjugate.        
It is noted that under the one-loop evolution of ${\rm Re}\hat E_F$ or ${\rm Im}\hat E_F$  will not be mixed with 
the imaginary- or real parts of three-parton FF's, respectively. The individual evolution of the real- and imaginary part 
of $\hat E_F$ can be read from the above equation directly.

\begin{figure}[hbt]
\begin{center}
\includegraphics[width=16cm]{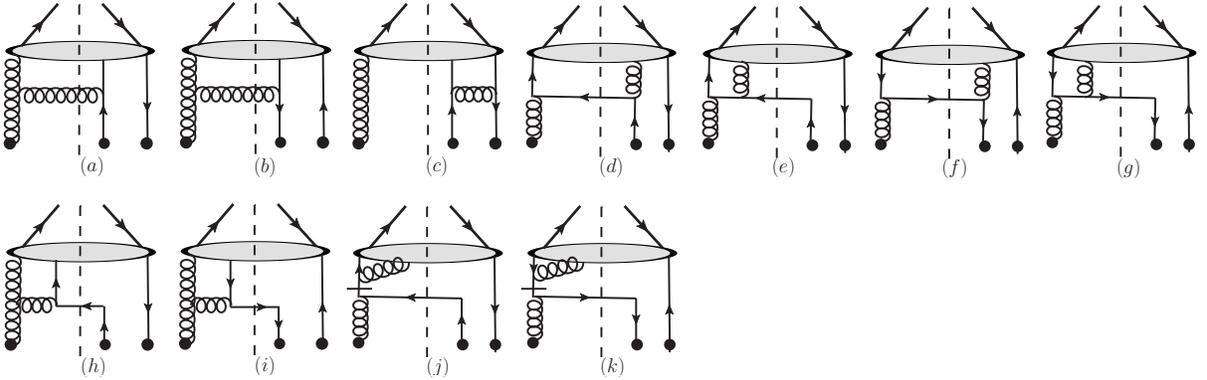}
\end{center}
\caption{Diagrams for the evolution of $\hat E_G$.  } 
\end{figure}

Summing the contributions from Fig.5 and the virtual part in Eq.(\ref{VIEF}), we obtain the evolution of $\hat E_G$: 
\begin{eqnarray} 
   \mu \frac{\partial \hat E_G (z_1,z_2) }{\partial \mu }  &=& \frac{\alpha_s }{\pi} \biggr \{ 
     C_F \biggr ( \frac{3}{2} +  \ln \frac{z_3}{z_1}  \biggr )  \hat E_G (z_1,z_2)
      + N_c \int_{z_2/z_1}^1 \frac{ d\xi_2}{(1-\xi_2)_+} \hat E_G (z_1,x_2)
\nonumber\\     
     && + \int_{0}^1 \frac{ d\xi_1}{(1-\xi_1)_+} \biggr ( N_c \hat E_G (x_1,x_{13})-\frac{1}{N_c} \hat E_G (x_1,z_2) \biggr )  
 +\frac{1}{2 N_c} \int^{z_1/z_2}_0 \frac{ d\xi_1}{1-\xi_1} \hat E_G (x_1,z_2)    
\nonumber\\   
  && +\frac{ N_c } {2 }  \int_{z_1}^\infty  \frac{d x_1}{x_1} \hat E_G(x_1,x_{13}  )  \frac{x_1^2 z_1 + x_1 z_2^2 - z_2^2 z_1}{x_1 ( x_1 (z_1-z_2) + z_1 z_2) } 
 +\frac{ 1 } { N_c  }  \int  \frac{d x_1}{x_1} \hat E_G(x_1, z_2 )
\biggr [ \theta (z_1-x_1)   
\nonumber\\   
  &&  + \theta (x_1-z_1) \frac{z_1}{x_1-z_2} \biggr ]
  -\frac{ N_c } {2 }  \int_{z_2}^{z_1}   \frac{d x_2}{x_2} \hat E_G(z_1,x_2 ) \biggr [ \frac{z_1^2}{z_2(x_2-z_1) } + \frac{x_2 (z_1-z_2) + z_2^2 }{x_2 z_2} \biggr ]    
\nonumber\\
 && -\frac{1  } {2 (N_c^2 -1)   } \biggr [  \int_{z_2}^{z_3}  \frac{d x_2}{x_2 } \biggr ( \hat E_F^*(z_3, x_2 ) + \hat E_{\bar F}^* (z_3, x_2 ) \biggr ) 
    \frac{z_3^2 (x_2-z_2)^2}{ x_2 z_2 (z_3-x_2)^2}     
\nonumber\\
 &&  +   \int  \frac{d x_1}{x_1}  \biggr ( \hat E_F(x_1, z_3 )  +\hat E_{\bar F} (x_1,z_3) \biggr ) \biggr (  
    \frac{z_3^2 (z_3+ z_2-2 x_1)}{ z_1 (x_1-z_3)^2}         
\nonumber\\
   &&      -  N_c^2 \frac{ x_1 z_1 z_3 + z_2^2 (2 z_1 + z_3)}{z_1 z_2 (x_1+z_1)}    \biggr ) \biggr ]\biggr \}.     
\label{EVOEG}         
\end{eqnarray}
Here, the evolution involves $\hat E_{\bar F}$. The evolution of $\hat E_{\bar F}$ can be derived from Eq.(\ref{EVOEF}) 
with the symmetry of charge-conjugation.  The results in Eq.(\ref{2PEVO}) and Eq.(\ref{EVOEF}, \ref{EVOEG}) are our main results. These results satisfy the relations in Eq.(\ref{RL2N}). This is an important check of our results.  

Our results in Eq.(\ref{EVOEF},\ref{EVOEG}) and the evolution of $\hat E_{\bar F}$ form a closed system of
differential-integral equations. The system contains six equations because $\hat E_{F, \bar F, G}$ are complex functions. 
In the large-$N_c$ limit, the system is simplified. Because of the normalization factor of color in the FF's definitions of 
Eq.(\ref{3PFF}), one can expect that all 3-parton FF's are at the same order of $N_c$ in the limit of $N_c \to \infty$. 
With this expectation one can easily find the evolutions in the limit from  Eq.(\ref{2PEVO}) and Eq.(\ref{EVOEF}). E.g., 
in the limit, the evolution of $\hat E_F$ becomes:
\begin{eqnarray} 
   \mu \frac{\partial \hat E_F (z_1,z_2) }{\partial \mu }  &=& \frac{\alpha_s N_c  }{2 \pi} \biggr \{ 
     \biggr (  \frac{3}{2} + \ln \frac{z_3^2}{z_1 z_2} \biggr ) \hat E_F (z_1,z_2)
     +  2\int_{z_2/z_1}^1 \frac{ d\xi_2}{(1-\xi_2)_+} \hat E_F(z_1,x_2) 
\nonumber\\
    &&  +  2\int_{0}^1  \frac{d\xi_1} {(1-\xi_1)_+}  \hat E_F (x_1,z_2)  
  -  \int^{z_1/z_2}_0 \frac{ d\xi_1}{1-\xi_1} \hat E_F(x_1,z_2) 
\nonumber\\      
   && + \frac{1}{z_3 } \int  \frac{d x_1}{x_1} \biggr \{\hat E_F(x_1,z_2)   \biggr [  
   \biggr (  \theta (x_1-z_1)  \frac{ z_2^2 (-x_1 z_2-x_1 z_3 +z_2 z_3)}{z_1 (x_1-z_2)^2}     
\nonumber\\
  && - \theta (z_1-x_1)\frac{z_2^2 (x_1-z_3) }{z_3 (x_1-z_2) }     \biggr ) 
  -  \frac{z_2^3}{z_1 (x_1-z_1) } \biggr ]   
 \nonumber\\  
 &&
 + \hat E_F^*(x_1,z_1) \frac{z_1z_2}{z_3(x_1-z_1)} \biggr [    (x_1+z_3)   +  \frac{ x_1^3}{(x_1-z_2)(x_1-z_1) }
  \biggr ] \biggr \}   
\nonumber\\
   && +  \int  \frac{d x_2}{x_2} \hat E_F(z_1,x_2)
      \theta(x_2-z_2) \biggr ( -\frac{z_2}{x_2} + \frac{z_1 ( z_1 x_2 + z_3 (z_1-x_2))}{z_3(x_2-z_1)^2 } \biggr ) \biggr \}
        + {\mathcal O }(N_c^0),  
\label{EVOEFNC}                       
\end{eqnarray}
from the above and the result for $\hat E_G$ we realize that there is no mixing between three-parton FF's under the evolution   in the large-$N_c$ limit. The system become six independent homogeneous equations.   
This observation has been first made in \cite{BeKu} for the real parts of three-parton FF's.           
From the results in Eq.(\ref{2PEVO}) two-parton FF's do not be mixed with $\hat E_G$ in the large-$N_c$ limit.

\par 
Our results about the evolution of $\hat e_I$ and imaginary parts of $\hat E_{F,\bar F, G}$ are new. The evolution of $\hat e_\partial$ has been derived in \cite{Kang}. The result agrees with our result of $\hat e_\partial$, except the contribution from 
${\rm Im}\hat E_G$, which is missing in \cite{Kang}.  The evolution of $\hat e$ and real parts of $\hat E_{F,\bar F, G}$ 
has been first given in \cite{BeKu}. But, we are unable to find a complete agreement with our results. 
\par 
         
\par 
To summarize: We have derived the evolutions of all chirality-odd twist-3 FF's at one-loop level. These evolutions 
satisfy the constraints from QCD equation of motion.  
From our results, the three-parton FF's will only be mixed with themselves under the evolutions, but not with two-parton FF's.  
In the large-$N_c$ limit, the mixing disappears and the evolutions of three-parton FF's  are governed by six homogeneous equations.  
With our results 
combined with evolutions existing in literature, the $Q$-dependence of single transverse-spin asymmetries can be now 
predicted completely.

\par\vskip40pt 
\noindent
{\bf Acknowledgments}
\par
The work is supported by National Natural 
Science Foundation of P.R. China(No.11275244, 11675241, 11605195). The partial support from the CAS center for excellence in particle 
physics(CCEPP) is acknowledged.

\par


\vskip30pt
\begin{thebibliography}{99}


	
\bibitem{FFR} A. Metz and A. Vossen,  Prog. Part. Nucl. Phys. 91 (2016) 136-202,   
e-Print: arXiv:1607.02521 [hep-ex]. 

\bibitem{Trans} R.L. Jaffe and X. Ji, Phys. Rev. Lett. {\bf 67} (1991) 552, Nucl. Phys. B375 (1992) 527. 

\bibitem{TW4DIS} R.K. Ellis, W. Furmanski and R. Petronzio, Nucl. Phys. B212 (1983) 29,  

\bibitem{JWQ} J.W. Qiu, Phys. Rev. D42 (1990) 30.

\bibitem{G2} X. Ji and J. Osborn, Nucl. Phys. B608 (2001) 235, e-Print: hep-ph/0102026.

\bibitem{BeKu} A.V. Belitsky and E.A. Kuraev, Nucl. Phys. B499 (1997) 301, e-Print:hep-ph/9611256.

\bibitem{Kang} Z.B. Kang, Phys. Rev. D83 (2011) 036006, e-Print:arXiv:1012.3419[hep-ph]. 

\bibitem{BBKT} I.I. Balitsky, V.M. Braun, Y. Koike and K. Tanaka,  Phys. Rev. Lett. {\bf 77} (1996) 3078, 
e-Print: hep-ph/9605439.

\bibitem{KoNi} Y. Koike and N. Nishiyama, Phys. Rev. D55 (1997) 3068, e-Print: hep-ph/9609207.

\bibitem{BD} A.V. Belitsky and D. Mueller,  Nucl. Phys. B503 (1997) 279,  e-Print: hep-ph/9702354.

\bibitem{Beli1} A.V. Belitsky, Phys. Lett.  B453  (1999)  59-72,  e-Print: hep-ph/9902361. 

\bibitem{KQ2} Z.-B Kang  and J.-W. Qiu, Phys.Lett. B713 (2012) 273-276, 
e-Print: arXiv:1205.1019 [hep-ph].  

   
\bibitem{MZGP} J.P. Ma, Q. Wang and G.P. Zhang, Phys. Lett. B718 (2013) 1358, e-Print:arXiv:1210.1006[hep-ph]. 

\bibitem{KQ} Z.-B. Kang and J.-W. Qiu, Phys. Rev. D79:016003,2009,
e-Print: arXiv:0811.3101 [hep-ph].

\bibitem{BMP}  V.M. Braun, A.N. Manashov and B. Pirnay, Phys. Rev. D80 (2009) 114002,
e-Print: arXiv:0909.3410 [hep-ph].

\bibitem{VoYu} W. Vogelsang and F. Yuan,  Phys.Rev. D79 (2009) 094010, 
e-Print: arXiv:0904.0410 [hep-ph].

\bibitem{ZhSc} A. Sch\"afer and J. Zhou, Phys. Rev. D85 (2012) 117501, e-Print: arXiv:1203.5293 [hep-ph].

\bibitem{MW} J.P. Ma and Q. Wang, Phys. Lett. B715 (2012) 157, e-Print:arXiv:1205.0611.    

\bibitem{JiBe} Y. Ji and A.V. Belitsky, Nucl. Phys. B894 (2015) 161, e-Print:arXiv:1405.2828[hep-ph]. 



\bibitem{JiFF} R.L. Jaffe and X. Ji, Phys. Rev. Lett. {\bf 71} (1993) 2547, e-Print:hep-ph/9307329, 
  X.-D. Ji, Phys. Rev. D49 (1994) 114,  
e-Print: hep-ph/9307235.

\bibitem{MPF} A. Metz and D. Pitonyak, Phys. Lett. B723 (2013) 365, e-Print: arXiv:1212.5037. 



\bibitem{EKT} H. Eguchi, Y. Koike and K. Tanaka, Nucl. Phys. B752 (2006) 1, e-Print:hep-ph/0604003. 



\bibitem{MeMe} S. Meissner and A. Metz, Phys. Rev. Lett. 102:172003,2009,  
e-Print: arXiv:0812.3783 [hep-ph]. 

\bibitem{PDFFF} J.C. Collins and D.E. Soper, Nucl. Phys. B194 (1982) 445.





\end{thebibliography}
\end{document}